\documentclass[twocolumn,showpacs,preprintnumbers,amsmath,amssymb,prb]{revtex4}
\usepackage{graphicx}% Include figure files
\usepackage{dcolumn}% Align table columns on decimal point
\usepackage{bm}% bold math

\begin{document}

\preprint{APS/123-QED}

\title{Tunneling and rattling in clathrate crystal}
\author{Terutaka Goto, Yuichi Nemoto, Takashi Yamaguchi, Mitsuhiro Akatsu and Tatsuya Yanagisawa}
\affiliation{Graduate School of Science and Technology, Niigata University, Niigata 950-2181, Japan}

\author{Osamu Suzuki and Hideaki Kitazawa}
\affiliation{National Institute for Materials Science, 1-2-1 Sengen, Tsukuba, Ibaraki 305-0003, Japan}
\date{\today}% It is always \today, today,

\begin{abstract}
We present tunneling and rattling motions of an off-center guest atom in cage referring to a clathrate crystal
La$_3$Pd$_{20}$Ge$_6$.
The elastic constant $C_{44}$ of La$_3$Pd$_{20}$Ge$_6$ shows a Debye-type dispersion around 20 K
obeying a relaxation time $\tau = \tau_0$exp$(E/k_{\rm{B}}T)$ with an attempt time $\tau_0 = 2.0\times10^{-12}$ sec
and an activation energy $E = 197$ K.
At low temperatures below 3 K down to 20 mK, the $C_{44}$ shows a softening of
$C_{44} = C_{44}^0(T-T_C^0)/(T-\mathit{\Theta})$ with $T_C^0 = -337.970$ mK and $\mathit{\Theta} = -338.044$ mK.
These facts are attributed to two different types of the off-center motions with $\Gamma_5$ symmetry in 4a-site cage of
La$_3$Pd$_{20}$Ge$_6$, a thermally activated rattling motion over the potential hill and a tunneling motion
through the potential hill at low temperatures.
\end{abstract}

\pacs{63.20.Pw,62.20.Dc,82.75.-z}% PACS, the Physics and Astronomy
\maketitle
\section{INTRODUCTION}
It has recently been found that filled skutterudites RM$_4$Sb$_{12}$ (R: La or Ce, M: Fe or Co) [1],
clathrate semiconductor compounds Sr$_8$Ga$_{16}$Ge$_{30}$ [2] and Eu$_8$Ga$_{16}$Ge$_{30}$ [3] show considerable reduction of
the thermal conductivity.
The rattling motion of an off-center guest atom over a potential hill in oversized cage in these clathrate compounds brings about reduction
in thermal phonon transport by a resonant phonon scattering,
which is favorable
for thermoelectric devices with profitable figure of merit [4].
Furthermore, the low-temperature behavior of thermal conductivity $\kappa(T) \sim T^2$ and
ultrasonic attenuation $\alpha(T) \sim T^3$ in Sr$_8$Ga$_{16}$Ge$_{30}$ [2] reminds us of the two-level system
due to tunneling motions in glasses with structural disorder [5-8]. It should be noted that the off-center guest atoms of clathrate compounds is located in cages forming an ideal periodic lattice.
The off-center tunneling in clathrate crystals, therefore, may be regarded as a new quantum degrees of freedom associated with a local Einstein phonon with a large density of state.

Quite recently, our group has found the frequency dependence in the elastic constant $C_{44}$ of
Ce$_3$Pd$_{20}$Ge$_6$ showing a ferroquadrupole ordering [9,10] and in ${(C_{11}-C_{12})/2}$ of
a filled skutterudite PrOs$_4$Sb$_{12}$ showing a heavy fermion superconductivity [11].
The ultrasonic dispersion in these compounds shows the fact that
a relaxation time for the rattling motion obeys thermal activation-type temperature dependence.
We have successfully projected out the off-center mode in cage, which couples to appropriate ultrasonic mode.
The phenomena associated with the tunneling motion are expected at low temperatures,
where the thermally activated rattling motion dies out completely.
However, the ferroquadrupole ordering at $T_Q = 1.25$ K in Ce$_3$Pd$_{20}$Ge$_6$ and
the superconductivity at $T_C = 1.85$ K in PrOs$_4$Sb$_{12}$ smear out the characteristic low-temperature properties due to tunneling motions. In the present paper, we have made ultrasonic measurements on La$_3$Pd$_{20}$Ge$_6$ with a cubic Cr$_{23}$C$_6$-type structure in order to examine the tunneling and rattling motions in the system being absent of the 4f-electron.

\section{Experiment}
An image furnace equipped with two ellipsoidal mirrors
was employed to grow a single crystal of La$_3$Pd$_{20}$Ge$_6$ with residual resistivity $\rho_0 = 2.7 \mu\Omega$cm.
The plane parallel surfaces with (100) or (110) orientation were determined by x-ray photograph.
The sound waves were generated and detected by the piezoelectric LiNbO$_3$ transducers bonded on the surfaces of specimen.
The change of sound velocity $v$ was detected by a phase comparator based on mixer technology.
The elastic constant $C = \rho v^2$ was estimated by the mass density $\rho = 10.179$ g/cm$^3$ of
La$_3$Pd$_{20}$Ge$_6$ with lattice parameter $a = 12.482$ {\AA} [11]. A $^3$He-evaporation refrigerator down to
450 mK and a $^3$He-$^4$He dilution refrigerator down to 20 mK were employed for low-temperature ultrasonic measurements.
Observation of the de Haas oscillation in elastic constants [12] ensures good quality of the present single crystal.

\section{Results and Discussions}
We show temperature dependence of elastic constants $C_{11}$, ${(C_{11}-C_{12})/2}$ and $C_{44}$ of La$_3$Pd$_{20}$Ge$_6$
in Fig. 1.
Here the longitudinal sound wave with frequencies of 70 MHz was employed for measurements of $C_{11}$ and
transverse one with 40 MHz for ${(C_{11}-C_{12})/2}$ and $C_{44}$.
With lowering temperature we have found a step-like increase around 20 K in $C_{44}$, while other modes of $C_{11}$
and ${(C_{11}-C_{12})/2}$ exhibit monotonous increase only.

\begin{figure}
\includegraphics[width=0.70\linewidth]{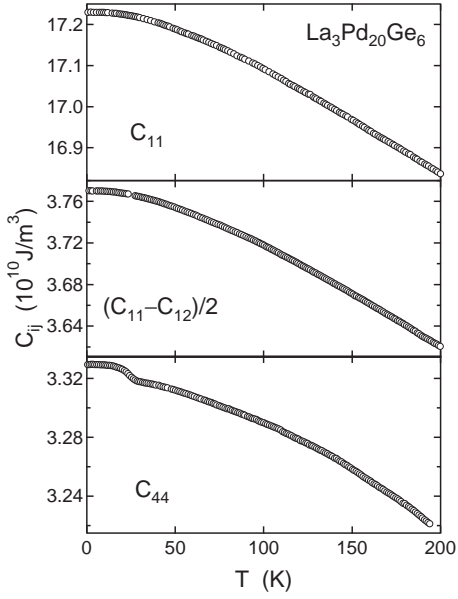}% Here is how to import EPS art
\caption{\label{ela}
Temperature dependence of elastic constants $C_{11}$, ${(C_{11}-C_{12})/2}$ and $C_{44}$ of La$_3$Pd$_{20}$Ge$_6$.
A step-like anomaly in $C_{44}$ around 20 K indicates the ultrasonic dispersion.
}
\end{figure}

The elastic constant $C_{44}$ of transverse acoustic mode in La$_3$Pd$_{20}$Ge$_6$ in Fig. 2(a) exhibits a marked dependence in frequencies for 9 MHz, 40 MHz and 150 MHz. This result indicates  the ultrasonic dispersion in the $C_{44}$ mode as similar as that in isomorphous compound Ce$_3$Pd$_{20}$Ge$_6$  [9].
These results show the fact that R$_3$Pd$_{20}$Ge$_6$ (R = La, Ce) exhibit commonly a thermally activated rattling motion of the off-center rare-earth ion.
As was noted in Fig. 1, the ultrasonic dispersion has been found only in  $C_{44}$ , but no sign of dispersion in ${(C_{11}-C_{12})/2}$ and  $C_{11}$. This result immediately means that the off-center motion has the $\Gamma_{5}$ symmetry. Consequently, the charge fluctuation due to the off-center motion couples to the acoustic transverse $C_{44}$ mode. 
On the other hand, a filled skutterudite PrOs$_4$Sb$_{12}$ of heavy fermion superconductor
exhibits ultrasonic dispersion in ${(C_{11}-C_{12})/2}$ indicating the $\Gamma_{23}$ symmetry of the off-center mode in cage [11]. It is worthwhile to emphasize that the symmetry transverse ultrasonic mode  is a useful probe to determine the symmetry of the off-center motion in cage.

The frequency dependence of the elastic constant in Fig. 2(a) is described by the Debye-type dispersion as

\begin{eqnarray}
C_{44}(\omega)&=&C_{44}(\infty)-\frac{C_{44}(\infty)-C_{44}(0)}{1+\omega^2\tau^2}.
\end{eqnarray}

The calculation of inset of Fig. 2(a) reproduces well the experimental results in Fig. 2(a).
Here $\omega$ is the angular frequency of the sound wave. The relaxation time $\tau$  of  the rattling motion in Fig. 2(b) revealed the thermally activated-type temperature dependence.
$C_{44}(0)$ and $C_{44}(\infty)$ are low- and high-frequency limits, respectively.
Arrows in Fig. 2(a) indicate the temperatures being satisfied with the resonant condition,
where the relaxation time $\tau$ coincides with the sound frequency $\omega$ as $\omega\tau = 1$.
The ultrasonic attenuation of the transverse $C_{44}$ mode, that is not presented here,
shows a peak at the temperatures satisfied with $\omega\tau = 1$. This resonance scattering of the ultrasonic $C_{44}$ wave by the rattling motion in cage may bring about a reduction of thermal conductivity by the phonon transport.

\begin{figure}
\includegraphics[width=0.8\linewidth]{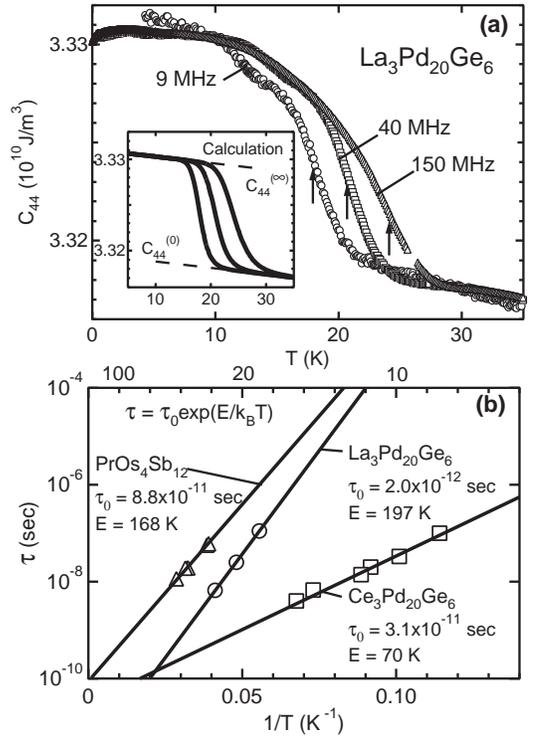}% Here is how to import EPS art
\caption{\label{dis}
(a) Frequency dependence of the elastic constant $C_{44}$ measured by the transverse ultrasonic waves with frequencies
of 9 MHz, 40 MHz and 150 MHz.
Arrows in (a) indicate the temperatures satisfying with the resonance condition of $\omega\tau$ = 1.
Inset in (a) represents a fit by the Debye-type dispersion of Eq. (1) for frequencies of 9 MHz, 40 MHz, and 150 MHz.
(b) shows the Arrhenius plot of relaxation time $\tau = \tau_0$exp$(E/k_{\rm{B}}T)$
in La$_3$Pd$_{20}$Ge$_6$ together with the results of Ce$_3$Pd$_{20}$Ge$_6$
and PrOs$_4$Sb$_{12}$ [9,11].
}
\end{figure}

It should be noted that the specified $\Gamma_5$ symmetry of the transverse $C_{44}$ wave may give rise to anisotropic phonon transport depending on the crystallographic orientation. The longitudinal $C_{11}$ mode as well as the transverse $C_{44}$ mode participates in the thermal phonon transport along the [100] direction. The later  $C_{44}$ mode is scattered by the rattling motin, while the former $C_{11}$ mode propagate transparently without any scattering by the rattling. In the case of thermal phonon transport along the [111] direction, the longitudinal  ${(C_{11}+2C_{12}+4C_{44})/3}$ as well as the transverse ${(C_{11}-2C_{12}+C_{44})/3}$ are scattered by the rattling,  because both modes contain the pure  $C_{44}$ mode  in part [13]. These anisotropic properties of acoustic phonon may bring about the anisotropy of the thermal conductivity by phonons. The investigation of the ratting by ultrasonic measurements would be profitable way to develop the thermoelectric device with high figure of merit.   

As one can see in an Arrhenius plot of Fig. 2(b),
the relaxation time $\tau$ of La$_3$Pd$_{20}$Ge$_6$ obeys the thermal activation-type temperature dependence
${\tau = \tau_0\rm{exp}(E/k_{\rm{B}}T)}$ with an attempt time ${\tau_0 = 2.0\times10^{-12}}$ sec
and an activation energy $E = 197$ K.
The relaxation times $\tau$ in Ce$_3$Pd$_{20}$Ge$_6$ with $\tau_0 = 3.1\times10^{-11}$ sec, $E = 70$ K [9]
and in PrOs$_4$Sb$_{12}$ with $\tau_0 = 8.8\times10^{-11}$ sec, $E = 168$ K [11]
are also presented in Fig. 2(b) for comparison.
These relaxations originate from the thermally assisted rattling motion of the off-center rare-earth ion
over a potential hill with height $E = 70 \sim 200$ K.
It is of importance to contrast with the charge fluctuation time of 4f electron
in inhomogeneous mixed valence compound Sm$_3$Te$_4$ showing the activation-type temperature dependence
with $\tau_0 = 2.5\times10^{-13}$ sec and $E = 1600$ K [14].
The relatively slow relaxation time $\tau_0 = 0.2 \sim 8.8\times10^{-11}$ sec as well as relatively small
activation energy $E = 70 \sim 200$ K in La$_3$Pd$_{20}$Ge$_6$, Ce$_3$Pd$_{20}$Ge$_6$ and PrOs$_4$Sb$_{12}$ indicate
the rattling motions of the off-center rare-earth ions with heavy masses.
The motion of rare-earth atom described by a harmonic oscillation of $\zeta(z) = (1/\pi z_0)^{1/2}$exp$(-z^2/2z_0^2)$
in cage spreads in space with a mean square displacement $z_0 = (1/2\pi)(h\tau_0/M)^{1/2}$ [15].
The attempt time $\tau_0 = 2.0\times10^{-12}$ sec and the mass $M = 139 m_p$ of La atom in La$_3$Pd$_{20}$Ge$_6$,
where $m_p$ is proton mass, lead to a mean square displacement $z_0 = 0.12$ {\AA} being comparable to
$z_0 = 0.48$ {\AA} of Ce$_3$Pd$_{20}$Ge$_6$, $z_0 = 0.79$ {\AA} of PrOs$_4$Sb$_{12}$ [9,11]. 
Furthermore neutron scattering on  Eu$_8$Ga$_{16}$Ge$_{30}$ shows $z_0 = 0.8$ {\AA} [16].

In order to examine low-temperature properties due to tunneling motions of La$_3$Pd$_{20}$Ge$_6$,
we have made ultrasonic measurements down to 20 mK. As shown in Fig. 3, we have observed a remarkable softening of $C_{44}$ with relative change 200 ppm below 3 K down to 20 mK by use of the dilution refrigerator. The ${(C_{11}-C_{12})/2}$ mode in inset of Fig. 3 shows
a softening of 20 ppm, which is much smaller than the 200 ppm softening of $C_{44}$.
The characteristic softening in $C_{44}$ of Fig. 3 in particular promises
the $\Gamma_5$ symmetry of the off-center tunneling motion in cage.
The tunneling frequency is estimated about 60 GHz corresponding to the onset temperature 3 K
in the elastic softening of Fig. 3.
The present ultrasonic measurements with frequencies $10 \sim 200$ MHz are able to observe
the static strain susceptibility for the off-center tunneling mode.
Magnetic fields up to 8 T do not make any appreciable change in the low-temperature softening of $C_{44}$.
This result, which is not presented here,  ensures that the softening of $C_{44}$ is intrinsic effect being independent of magnetic impurity effects.
\begin{figure}
\includegraphics[width=0.8\linewidth]{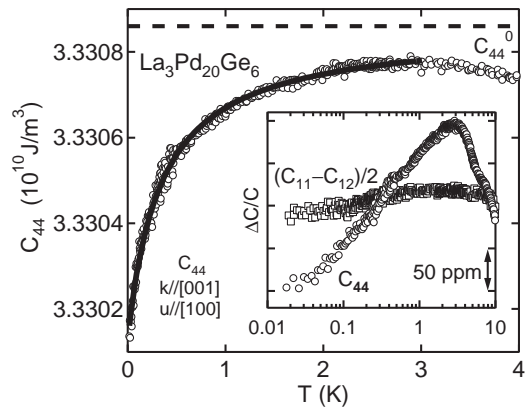}% Here is how to import EPS art
\caption{\label{sof}
The low-temperature softening in elastic constant $C_{44}$ of La$_3$Pd$_{20}$Ge$_6$ below 3 K down to 20 mK.
Solid line is a fit by $C_{44} = C_{44}^0(T-T_C^0)/(T-\mathit\Theta)$ with the parameters in the text.
Inset presents low-temperature behavior of $C_{44}$ and ${(C_{11}-C_{12})/2}$.
}
\end{figure}

In the present clathrate compound La$_3$Pd$_{20}$Ge$_6$ with space group symmetry $Fm\bar{3}m$,
there are two different types of cages containing La ion. The unit cell of La$_3$Pd$_{20}$Ge$_6$
with a lattice parameter $a = 12.482$ {\AA} consists of four molecular units with 116 atoms.
The 4a-site cage with La1 atom consisting of twelve Pd-atoms with distances $d_{\rm{La1-Pd2}} = 3.084$ {\AA}
and six Ge-atoms with $d_{\rm{La1-Ge}} = 3.350$ {\AA} forms a face center cubic lattice.
The 8c-site cage with La2 atom consisting sixteen Pd atoms with $d_{\rm{La2-Pd1}} = 2.884$ {\AA}
and $d_{\rm{La2-Pd2}} = 3.392$ {\AA} forms a simple cubic lattice [10].
The trivalent La ion with radii 1.88 {\AA} in former oversized 4a-site cage may favor the rattling motion
over off-center positions.
Actually neutron scatterings on R$_3$Pd$_{20}$Ge$_6$ (R = Ce, Pr, Nd) observed well-defined
crystalline electric field excitations of 8c site, while they detected obscure peak only for 4a site [17].
These facts are consistent with the off-center motion of the rare-earth ion in the 4a site cage of R$_3$Pd$_{20}$Ge$_6$.
Because the absence of Pd- or Ge-atom on the three-fold [111] axis in the 4a-site cage,
the potential for La1 ion may possess off-center minima at
$r_1 = (a,a,a)$, $r_ 2 = (-a,-a,a)$, $r_3 = (-a,a,-a)$, $r_4 = (a,-a,-a)$,
$r_5 = (a,a,-a)$, $r_6 = (-a,-a,-a)$, $r_7 = (a,-a,a)$, $r_8 = (-a,a,a)$
with $a \sim z_0/2 = 0.06$ {\AA}.
Here $r_i (i = 1,2,...,8)$ denotes distance from the center of the 4a-site cage.

It is useful to project out the irreducible representation for the off-center modes consisting
of the atomic densities $\rho_i = \rho(r_i)$ at the off-center positions $r_i$
in 4a-site cage with the site symmetry O$_{\rm{h}}$ [9, 18, 19].
The off-center motion over the eight off-center positions is reduced into direct sum
$\Gamma_1 \oplus \Gamma_2 \oplus \Gamma_4 \oplus \Gamma_5$.
Because the $C_{44}$ mode with $\Gamma_5$ symmetry in La$_3$Pd$_{20}$Ge$_6$ shows the ultrasonic dispersion around 20 K
in Fig. 2(a) and the appreciable low-temperature softening below 3 K in Fig. 3, the $\Gamma_5$ off-center mode
$\rho_{\Gamma 5,yz}=\rho_1-\rho_2-\rho_3+\rho_4-\rho_5+\rho_6-\rho_7+\rho_8$,
$\rho_{\Gamma 5,zx}=\rho_1-\rho_2+\rho_3-\rho_4-\rho_5+\rho_6+\rho_7-\rho_8$,
$\rho_{\Gamma 5,xy}=\rho_1+\rho_2-\rho_3-\rho_4+\rho_5+\rho_6-\rho_7-\rho_8$
is relevant for the ground state in the present system.
The $\Gamma_5$ mode consists of anisotropic fractional atomic state over eight off-center positions
$r_i (i =1, 2,...,8)$.
The $\rho_{\Gamma 5,xy}$ mode of Fig. 4, for instance, has fraction 1/4 at $r_1, r_2, r_5, r_6$ positions and null at $r_3, r_4, r_7, r_8$.
Other modes with $\Gamma_1, \Gamma_2$ or $\Gamma_4$ symmetry probably correspond to exited states and
hence they are disregarded here. The Schr\"{o}dinger equation for the off-center atom in multi potential wells is a faithful way to analyze the off-center tunneling state [20,21].
The present group theoretical analysis is a proper way instead of rigorous treatment. 

Because the La ion is trivalent, the charge density associated with the $\Gamma_5$ off-center mode in the 4a-site cage
is described as $\Delta\rho = Q_{yz}\rho_{\Gamma 5,yz} + Q_{zx}\rho_{\Gamma 5,zx} + Q_{xy}\rho_{\Gamma 5,xy}$ [18,19].
Here $Q_{yz}, Q_{zx}$ and $Q_{xy}$ denote the normal coordinate describing the $\Gamma_5$ off-center mode in Fig. 4. It is notable that the hexadecapole as well as quadrupole are relevant for the charge distribution of the  $\Gamma_5$ off-center mode. 
As the temperature is lowered, the phase transition $\Delta \rho \neq 0$ due to the condensation
of the symmetry breaking $\Gamma_5$ mode would be expected in principle.
We refer Yb$_4$As$_3$ to the phase transition due to charge ordering [18].
The present system, however, no phase transition has been observed down to 20 mK.
Consequently, the tunneling motion of the off-center La-ion through the potential hill
in keeping the O$_{\rm{h}}$ site symmetry is relevant even at low temperatures.

The interaction of the $\Gamma_5$ off-center tunneling mode in cage to the elastic strain
$\varepsilon_{yz}, \varepsilon_{zx}, \varepsilon_{xy}$ of the $C_{44}$ mode is described as [22-24]

\begin{eqnarray}
H_{QS} = -g_{\Gamma_5}\sum_i\{Q_{yz}(i)\varepsilon_{yz} + Q_{zx}(i)\varepsilon_{zx} \nonumber \\ + Q_{xy}(i)\varepsilon_{xy}\}.
\end{eqnarray}

Here $g_{\Gamma 5}$ is a coupling constant and $\Sigma_i$ means summation over cages at site $i$.
The inter-cage coupling of the  $\Gamma_5$ off-center tunneling mode accompanying the charge fluctuation  is introduced as

\begin{eqnarray}
H_{QQ} = -g_{\Gamma_5}'\sum_i\{\langle Q_{yz} \rangle Q_{yz}(i) + \langle Q_{zx} \rangle Q_{zx}(i) \nonumber \\ + \langle Q_{xy} \rangle Q_{xy}(i)\}.
\end{eqnarray}

Here $g_{\Gamma 5}'$ is a coupling constant and $\langle Q_{ij} \rangle (i,j = x, y, z)$ denotes a mean field.
The low-temperature elastic softening in $C_{44}$ below 3 K in Fig. 3 is described by

\begin{eqnarray}
C_{44}&=&C_{44}^0 -\frac{Ng_{\Gamma_5}^2\chi _{\Gamma_5}(T)}{1-g'_{\Gamma_5} \chi_{\Gamma_5} (T)}.
\end{eqnarray}

Here $N$ is the number of cages in unit volume.
The strain susceptibility for the $\Gamma_5$ off-center mode in cage is written as
$g_{\Gamma 5}^2\chi_{\Gamma 5}(T) = C_Q/T = \delta^2/T$.
Here is $\delta = (C_Q)^{1/2}$ is a deformation coupling energy for the $\Gamma_5$ ground state.
Consequently, one obtains $C_{44} = C_{44}^0(T-T_C^0)/(T-\mathit\Theta)$
with $\mathit\Theta = (g_{\Gamma 5}'/g_{\Gamma 5}^2)C_Q$ and $T_C^0 = NC_Q/C_{44}^0 + \mathit\Theta$.
The solid line in Fig. 3 is a fit by the characteristic temperatures $\mathit{\Theta} = -338.044$ mK, $T_C^0 = -337.970$ mK
and a back ground $C_{44}^0 = 3.33085\times10^{10}$ J/m$^3$.
The low-temperature softening of $C_{44}$ proportional to reciprocal temperature in La$_3$Pd$_{20}$Ge$_6$
is attributed to the $\Gamma_5$ tunneling mode with the symmetry breaking character in the present system.
This is strikingly different from minima in elastic constants of NaCl, which has an off-center tunneling motion
with $\Gamma_1$ singlet ground state of impurity OH ion [25,26].

The value of $NC_Q/C_{44}^0 = T_C^0-\mathit{\Theta} = 0.074$ mK corresponds to the energy gain due to the coupling of
the $\Gamma_5$ off-center mode to the elastic strain as similar as the Jahn-Teller energy
in d- or f-electron system [22-24].
Employing the number of the 4a site in unit volume, $N = 2.057\times10^{21}$ cm$^{-3}$,
one obtains the deformation coupling energy $\delta = 9.3$ K
corresponding to the energy shift for the $\Gamma_5$ mode by unit external strain.
It is remarkable that the deformation coupling of about 10 K for the present tunneling mode is order of
magnitude smaller than that of 100 K in 4f-electron system of rare-earth compounds and of 1000 K in 3d-electron system of
transition metal compounds.
This means that the charge fluctuation due to the tunneling motion is well screened by the cage.
The negative value of $\mathit{\Theta} = -338.044$ mK indicates the antiferro-type inter-cage interaction
of the $\Gamma_5$ mode among the 4a-site cages with nearest neighbor distance 8.8 {\AA}.
\begin{figure}
\includegraphics[width=0.52\linewidth]{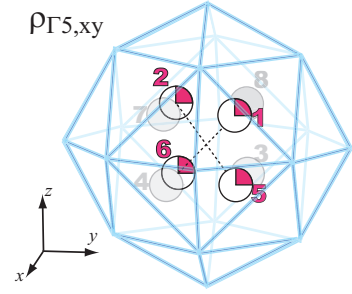}% Here is how to import EPS art
\caption{\label{rat}
Schematic view of the $\Gamma_5$ off-center mode
$\rho_{\Gamma_5,xy}=\rho_{1}+\rho_{2}-\rho_{3}-\rho_{4}+\rho_{5}+\rho_{6}-\rho_{7}-\rho_{8}$
possessing fractional states of La1-ion over eight off-center potential minima along three fold [111] axis
in 4a-site cage of La$_3$Pd$_{20}$Ge$_6$.
}
\end{figure}

The tunneling motion in the two-level system of many structural glasses [5,27] provides the principal concept to elucidate the off-center tunneling motion of the clathrate crystals.
The two-level system in ramdom potential of structural glasses losses its symmetry property.
This is strikingly different from the specific $\Gamma_5$ symmetry of the off-center tunneling motion in periodically arrayed cages of the clathrate crystal La$_3$Pd$_{20}$Ge$_6$. As was already shown, the off-center tunneling in the present La$_3$Pd$_{20}$Ge$_6$ has small antiferro- type inter-cage interaction $\mathit{\Theta} = -338.044$ mK. Therefore, the collective motion of the off-center  tunneling in periodically arrayed cages is expected to show a flat dispersion being mostly independent of the wave vector. The triply degenerate  $\Gamma_5$ tunneling state may be regarded as  a local Einstein phonon with large density of state.

It is expected that  the off-center tunneling motion in cage of La$_3$Pd$_{20}$Ge$_6$ interacts to the conduction electron consisting
of the 4d orbit of Pd and 4p orbit of Ge of cage. This interaction of the local Einstein phonon to the electrons may play a crucial role to bring about the small activation energy
$E = 197$ K and the appreciable mean square displacement $z_0 = 0.12$ {\AA}.
We noted many theoretical investigations showing that in a strong electron-phonon coupling regime, the renormalized potential for a guest atom may favor off-center positions and reduction of the excitation energy [15, 28-29].
The screening of the off-center tunneling mode, namely local Einstein phonon is mapped to the multi-channel Kondo model of the non-Kramers doublet in f-electron systems [29].

In the case of the heavy Fermion superconductor PrOs$_4$Sb$_{12}$, the ultrasonic dispersion has been found in the ${(C_{11}-C_{12})/2}$ mode [11]. The doubly degenerate $\Gamma_{23}$ tunneling state accompanying the charge fluctuation may couple with the quadrupole fluctuation of the  $\Gamma_1$- $\Gamma_4^{(2)}$ pseudo quartet CEF state as well as the conduction band [31]. This interaction of the  local phonon to the electrons is of crucial importance for the formation of the heavy Fermion state and its Cooper pair of the superconductivity. The tunneling state with doubly or triply degeneracy in cage of clathrate compounds may regarded as a new quantum degrees of freedom  leading to exotic low-temperature properties concerning the symmetry breaking phase transition, superconductivity, and the multi-channel Kondo effect.

\section{Conclusion}
In conclusion, the present ultrasonic measurements on the clathrate compound La$_3$Pd$_{20}$Ge$_6$ revealed the dispersion in the transverse $C_{44}$ mode around 20 K due to the thermally assisted rattling motion over the potential hill. We have also found the appreciable elastic softening of $C_{44}$ below 3 K down to 20 mK
attributed to the off-center tunneling motion through the potential hill. The off-center motion in the present clathrate crystal La$_3$Pd$_{20}$Ge$_6$ is an ideal local Einstein oscillator with triple degeneracy  in periodically arrayed  cages. The clarification of the exotic properties of the local Einstein phonon being strongly coupled to electrons in metallic clathrate compounds is an interesting  issue of strongly correlated physics in future. 

\begin{acknowledgments}
We thank M.Goda, H.Iyetomi, K.Miyake and Y.$\bar{\rm{O}}$no for stimulated discussions.
The present work was supported by a Grant-in-Aid for Scientific Research Priority Area (No.15072206)
of the Ministry of Education, Culture, Sports, Science and Technology of Japan.
\end{acknowledgments}

\end{document}